\documentclass{moriond}

\def\Journal#1#2#3#4{{#1} {\bf #2}, #3 (#4)}

\def\PLB{{\em Phys. Lett.}  B}
\def\PRL{\em Phys. Rev. Lett.}
\def\PRD{{\em Phys. Rev.} D}
\def\PRC{{\em Phys. Rev.} C}

\def\EPJC{{\em Eur. Phys. J.} C}
\def\PhysRep{{\em Phys. Rep.}}
\def\JINST{{\em JINST}}

\def\be{\begin{equation}}
\def\ee{\end{equation}}
\def\bea{\begin{eqnarray}}
\def\eea{\end{eqnarray}}

\newcommand{\Photo}{}

\begin{document}
\vspace*{4cm}
\title{Coherent J/$\psi$ photoproduction and polarization in peripheral Pb--Pb collisions with ALICE}

\author{ L. Massacrier for the ALICE Collaboration }

\address{Universit\'e Paris-Saclay, CNRS/IN2P3, IJCLab, Orsay, France}

\maketitle\abstracts{
Photonuclear reactions are induced by the strong electromagnetic field generated by ultrarelativistic heavy ions. These processes have been extensively studied in ultraperipheral collisions. In recent years, the observation of coherent J/$\psi$ photoproduction has been claimed in nucleus--nucleus (A--A) collisions with nuclear overlap, based on the measurement of an excess in the very low transverse momentum ($p_{\rm T}$) J/$\psi$ yield. Such quarkonium measurements can help constraining the nuclear gluon distribution at low Bjorken-$x$ and can shed light on the theory behind photon induced reactions in A--A collisions with nuclear overlap. In order to confirm the photoproduction origin of the very low-$p_{\rm T}$ J/$\psi$ yield excess, polarization measurement is a golden observable. It is indeed expected that the produced quarkonium would keep the polarization of the incoming photon due to $s$-channel helicity conservation. ALICE can measure inclusive and exclusive quarkonium production down to zero transverse momentum, at forward rapidity (2.5 $< y <$ 4) and midrapidity ($|y| <$ 0.9). In this contribution, we will report on the new preliminary measurement of the $y$-differential cross section and the first polarization analysis at LHC of coherently photoproduced J/$\psi$ in peripheral Pb--Pb collisions. Both measurements are conducted at forward rapidity in the dimuon decay channel.
}


Photonuclear reactions can be produced by the strong electromagnetic field generated by ultra-relativistic heavy ions, and this field can be treated as a flux of quasi-real photons. The photon can oscillate into a quark-antiquark pair (a color dipole), that can probe the gluon distribution in the target nucleus via the exchange of two gluons (leading order approximation), before recombining to form a vector meson (VM). The photoproduction of charmonium (a bound state of a charm and anticharm quarks) in A--A collisions is therefore a tool sensitive to nuclear gluon distributions down to very low Bjorken-$x$, in the range $\sim$ 10$^{-2}$ to 10$^{-5}$ at LHC energies, ie. a region where those distributions are still poorly constrained \cite{Eskola}. Such photonuclear reactions have been extensively studied in ultraperipheral nucleus--nucleus collisions (UPC) \cite{Baltz}, in which the impact parameter of the collision is larger than twice the nuclear radius. Charmonia can be photoproduced on nuclei either coherently or incoherently. In the coherent interaction, the photon couples to the target nucleus as a whole, leaving it intact and producing a VM of very low transverse momentum ($p_{\rm T} \sim$ 60 MeV/$c$). In the incoherent case, the photon couples to a single nucleon, usually leading to the breaking of the target nuclei. In this case, the VM is produced with larger momenta ($p_{\rm T} \sim$ 500 MeV/$c$). The ALICE Collaboration reported in 2016 the first observation of a very low-$p_{\rm T}$ excess in the yield of J/$\psi$ with respect to expectations from hadroproduction, in Pb--Pb collisions \cite{ALICE1}. Coherent photoproduction of J/$\psi$ in nucleus--nucleus collisions with nuclear overlap was proposed as a plausible mechanism to explain this excess. Further measurements by STAR \cite{STAR}, LHCb \cite{LHCb} and ALICE \cite{ALICE2} of the phenomena corroborate this hypothesis. The coexistence of coherent photoproduction together with a hadronic interaction raises several theoretical questions. In particular how can the coherence condition survive if the target nucleus is broken during the hadronic interaction? Do only the spectator nucleons participate in the coherent process? Can the photoproduced J/$\psi$ interact with the quark-gluon plasma produced in A--A collisions? Recent theoretical developments which address these points are all based on UPCs calculations extended to A--A collisions with nuclear overlap~\cite{Cepila,Ducati,Zha}. In such framework, the photoproduction cross section of a VM is computed as the product of a quasi-real photon flux with a photonuclear cross section corresponding to the $\gamma$A~$\rightarrow$~VM+A interaction. In order to account for the nuclear overlap, models usually consider an effective photon flux to take into account the geometrical constraints of a given impact parameter range. In few models, an effective photonuclear cross section is also considered to account for the effective size of nuclear fragments participating in the coherent process. Precise rapidity-differential cross section measurements of coherently photoproduced J/$\psi$ in peripheral Pb--Pb collisions should permit to further constrain those newly developed models. In addition, the polarization measurement of photoproduced J/$\psi$ is expected to be a golden observable to probe the production mechanism at play. Indeed, if the J/$\psi$ is photoproduced, it is expected to keep the transverse polarization of the incoming photon due to $s$-channel helicity conservation \cite{Gilman}, while on the other hand, hadronic J/psi polarization measurements in Pb-Pb indicate only small polarization in the helicity frame ($\lambda_{\theta} \sim$ 0.2 at low $p_{\rm T}$) \cite{ALICE3}. 


The ALICE detector is described in detail in Ref \cite{ALICE4}. In the analyses discussed below, the J/$\psi$ is reconstructed at forward rapidity (2.5 $< y <$ 4) in the dimuon decay channel, down to $p_{\rm T}$~=~0, with the forward muon spectrometer. The spectrometer consists of a front absorber of composite material, a tracking system, a muon filter made of iron and a triggering system. The Silicon Pixel Detector, which composes the two innermost layers of the central barrel Inner Tracking System, is used to reconstruct the interaction vertex. The V0 scintillators, located on both sides of the interaction point, are used for triggering, beam-gas background rejection and the determination of the collision centrality \cite{ALICE5}. The ZERO Degree Calorimeters placed on both side of the interaction point, measure spectator protons and neutrons. A minimum energy deposition in the two neutron  calorimeters, corresponding to the expected signal from one neutron spectator, is required to suppress the background induced by electromagnetic dissociation processes, in combination with V0 and ZDC timing information. The data sample analyzed corresponds to the total Run 2 statistics of Pb--Pb collisions at $\sqrt{s_{\rm NN}}$ = 5.02 TeV. Selected events satisfy the following requirements: the detection of two opposite sign muons in the muon trigger with a $p_{\rm T}$ above the trigger threshold of 1 GeV/$c$ for each track, and a coincidence with the minimum bias trigger. The minimum bias trigger requires the coincidence of a signal in both arrays of the V0 detector. The integrated luminosity used in the analyses below is 756~$\pm$~19~$\mu b^{-1}$. The results are all provided for the centrality range 70-90$\%$, corresponding to peripheral collisions.

\begin{figure}[!htpb]
\begin{minipage}{0.5\linewidth}
\centerline{\includegraphics[width=1.\linewidth]{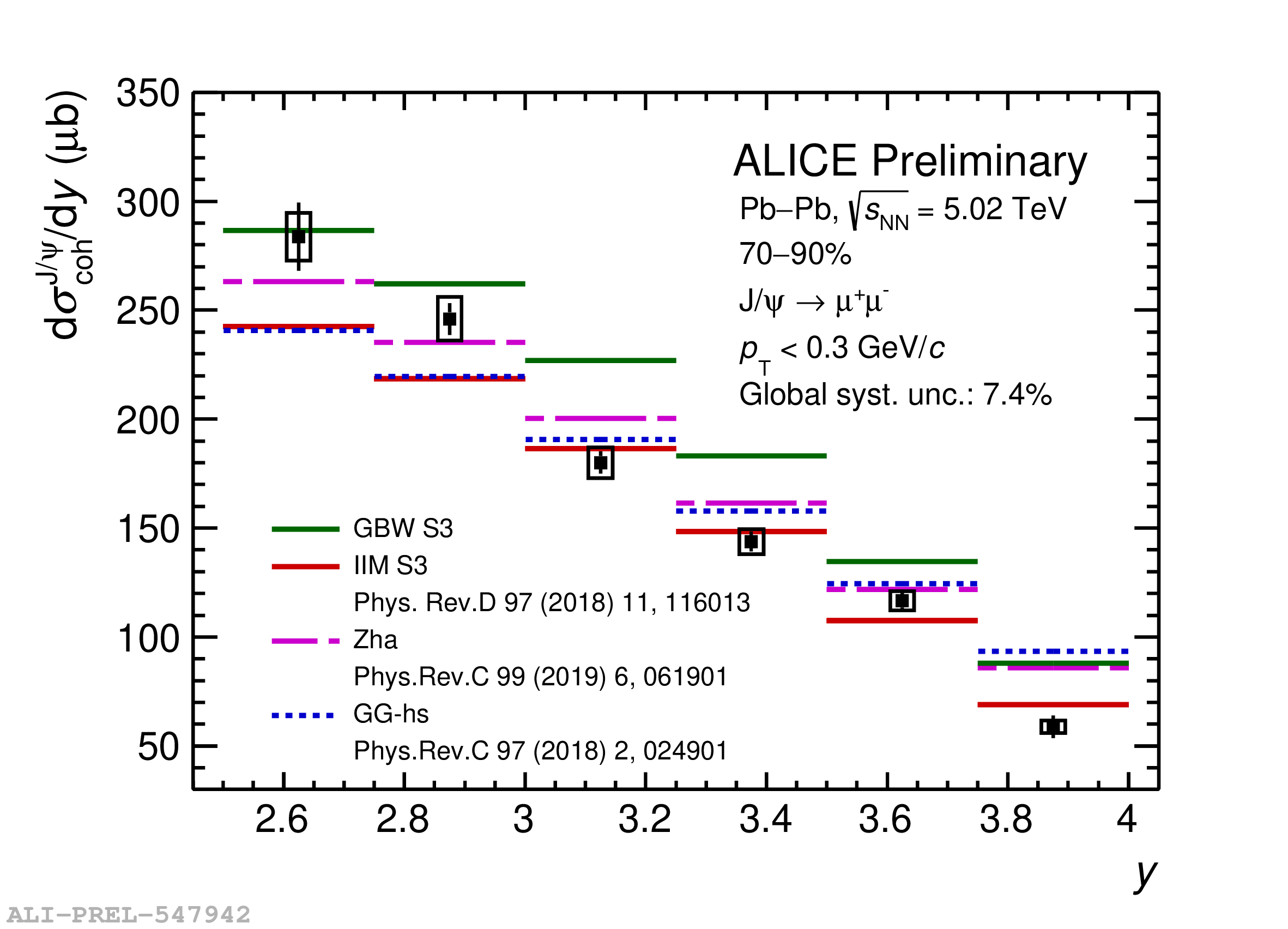}}
\end{minipage}
\hfill
\begin{minipage}{0.5\linewidth}
\centerline{\includegraphics[width=1.\linewidth]{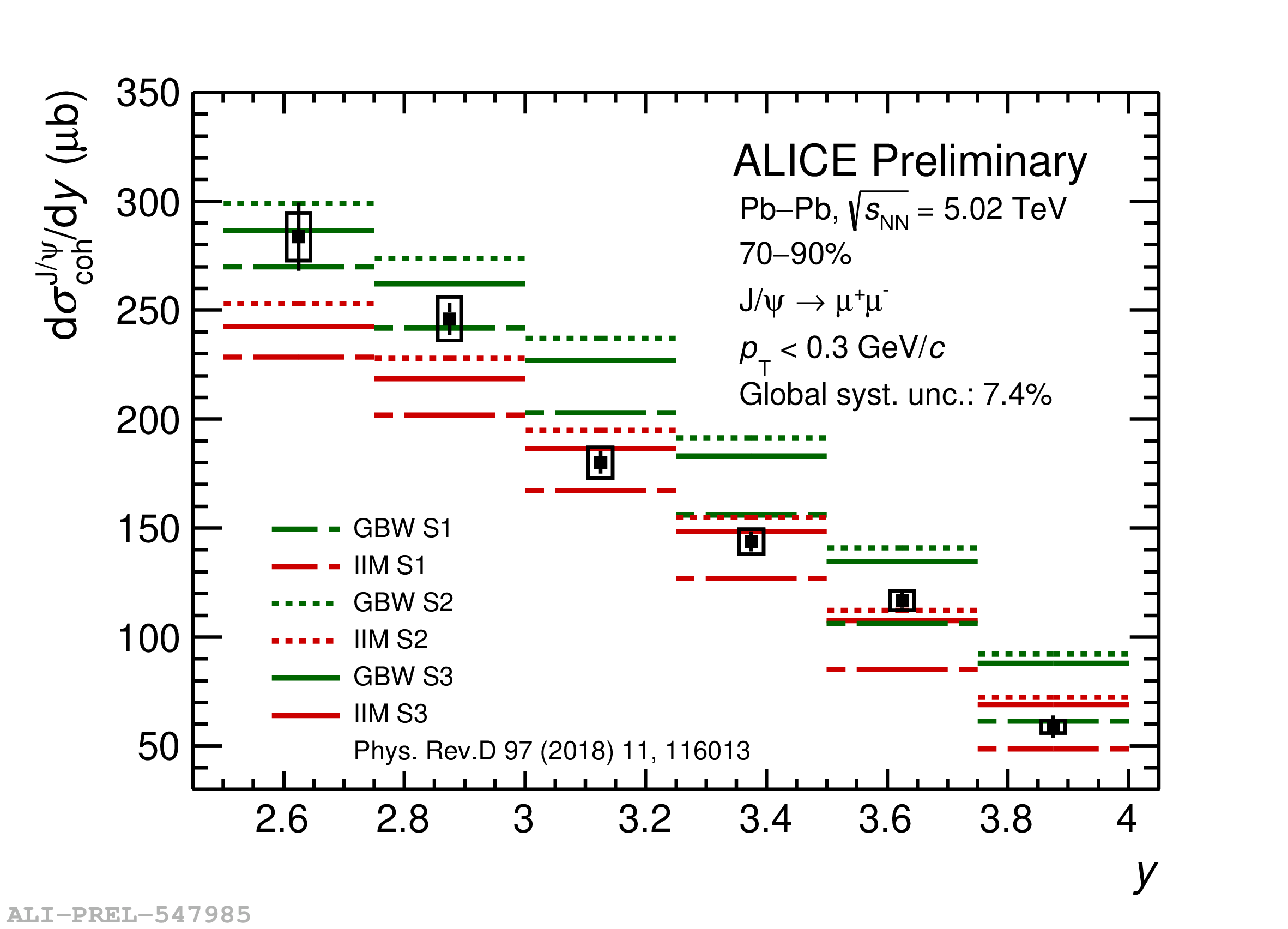}}
\end{minipage}
\hfill
\caption[]{Rapidity-differential J/$\psi$ photoproduction cross section measured by ALICE in peripheral Pb-Pb collisions at $\sqrt{s_{\rm NN}}$ = 5.02 TeV, for $p_{\rm T} <$ 0.3 GeV/$c$ and 2.5 $< y <$ 4. Data are compared to the GBW/IIM S3 model \cite{Ducati} (solid line), the Zha model \cite{Zha} (dashed-dotted line), the GG-hs model \cite{Cepila} (dotted line) in the left panel, and to the GBW/IIM S1 (dashed-dotted line), S2 (dotted line), S3  model \cite{Ducati} (solid line) in the right panel.}
\label{fig12}
\end{figure}

The forward rapidity-differential J/$\psi$ photoproduction cross section measured by ALICE, in the centrality range 70-90$\%$, for $p_{\rm T} <$ 0.3 GeV/$c$, in Pb--Pb collisions at $\sqrt{s_{\rm NN}}$ = 5.02 TeV, is shown in Fig. \ref{fig12}. The cross section exhibits a smooth decrease towards forward rapidities. Data are compared to several models developed for UPC physics and extendend to peripheral A--A collisions. The GG-hs \cite{Cepila} is a hot-spot based model, where the hot spots represent subnucleonic degrees of freedom. The calculation uses Glauber-Gribov formalism to compute the dipole-nucleus cross section based on the dipole-proton cross section. The photon flux is computed in a similar way as in the UPC case, but the integral is limited to the impact parameter range corresponding to the centrality 70-90$\%$. The IIM/GBW model \cite{Ducati} is based on a vector dominance model and the color dipole formalism. The dipole-cross section is computed either within the GBW light cone dipole formalism \cite{GBW} or the IIM color glass condensate based approach \cite{IIM}. Three scenarios are considered in this model. In S1, no relevant modifications of the photon flux and photonuclear cross section with respect to UPC are performed. In S2, an effective photon flux is used, in which the UPC flux is modified in order to consider only the photons that reach the geometrical region of the target nucleus outside of the overlap region. In S3, on top of the effective photon flux from S2, an effective photonuclear cross section is introduced, in which the overlap region between the two nuclei is excluded. The Zha model \cite{Zha} is based on a double-slit framework at the fermi scale. The hadronic interaction effects on the photon flux are considered. Nuclear shadowing and coherent length effects are also accounted for. All models are qualitatively able to describe the magnitude of the measured J/$\psi$ photoproduction cross section, but globally fail at describing quantitatively its rapidity dependence. Similar observation was also made in the UPC case where photoproduction models are not able to describe the measured cross section in the full rapidity range \cite{ALICE6}. 

\begin{figure}[!htpb]
\begin{minipage}{0.5\linewidth}
\centerline{\includegraphics[width=0.9\linewidth]{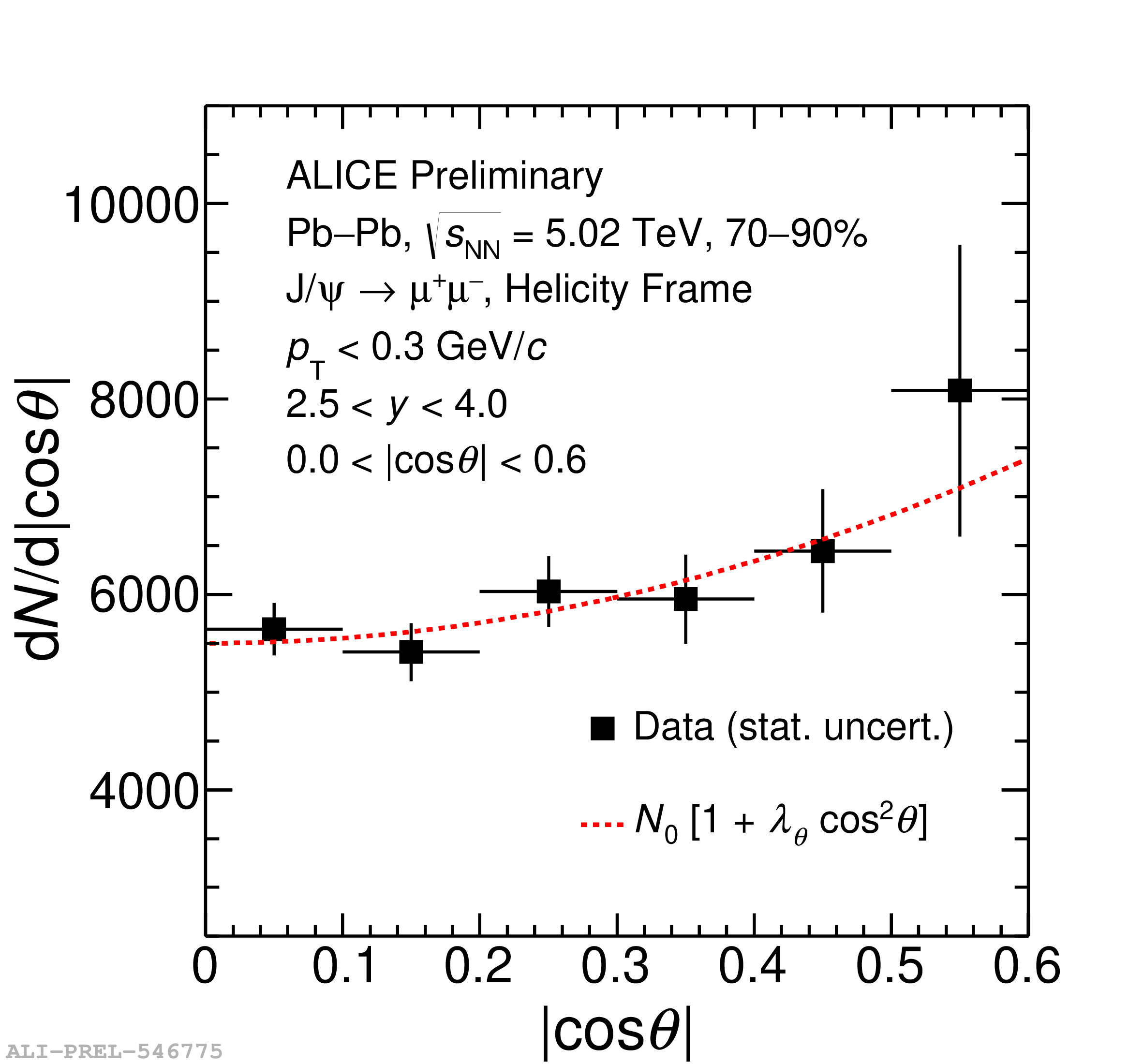}}
\end{minipage}
\hfill
\begin{minipage}{0.5\linewidth}
\centerline{\includegraphics[width=0.9\linewidth]{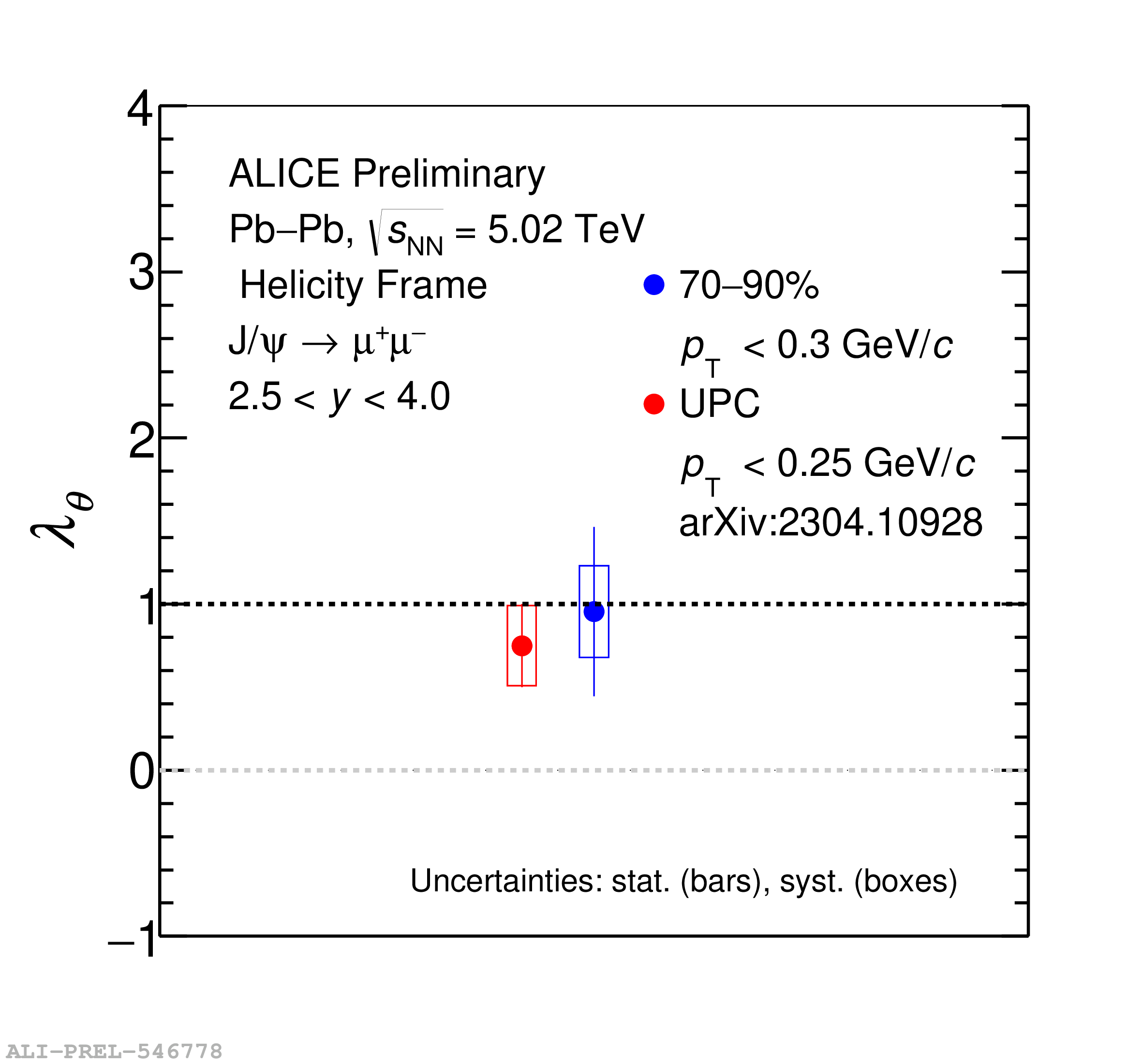}}
\end{minipage}
\hfill
\caption[]{Inclusive J/$\psi$ corrected yield as a function of $|$cos$\theta|$ (left) and extracted $\lambda_{\theta}$ parameter (right), measured by ALICE in the helicity frame, in peripheral Pb-Pb collisions at $\sqrt{s_{\rm NN}}$ = 5.02 TeV, for $p_{\rm T} <$~0.3~ GeV/$c$ and 2.5~$< y <$4. In the right panel, the peripheral Pb--Pb results are compared to similar measurement performed in UPC collisions \cite{ALICE7} at the same c.o.m energy, in the same reference frame and rapidity range, for $p_{\rm T} <$ 0.25 GeV/$c$.}
\label{fig34}
\end{figure}

Fig \ref{fig12} right, shows that the impact of the nuclear overlap in the centrality range 70-90$\%$ on the measured cross section (the difference between scenarios S1/S2/S3), is as large as the difference bewteen the model computations using GBW or IIM formalisms, which differ only by the assumptions made to compute the dipole-proton cross section. In order to interpret the peripheral Pb--Pb results and the role of the nuclear overlap on the J/$\psi$ photoproduction cross section, model inputs will require to be further constrained with UPC data in the first place. In addition, it would be interesting to extend the $y$-differential J/$\psi$ photoproduction cross section measurement to more central events, where the impact of the nuclear overlap is expected to be stronger. Fig. \ref{fig34} (left) shows the inclusive J/$\psi$ corrected yield as a function of $|\rm cos\theta|$ measured by ALICE in the helicity frame, in peripheral Pb--Pb collisions, for $p_{\rm T} <$0.3~GeV/$c$ and 2.5~$< y <$~4. The polar angle $\theta$ is defined by the direction of one of the J/$\psi$ decay daughter with respect to the chosen polarization axis. In the helicity frame, the polarization axis is given by the flight direction of the quarkonium in the center-of-mass (c.o.m) of the colliding beams. The J/$\psi$ $|\rm cos \theta|$ distribution is fitted with the following expression~:~W($|\rm cos \theta|$)~=~ $N_{0}$~(1~+~$\lambda_{\theta}$~cos$^{2}\theta$), where $\lambda_{\theta}$ is one of the three polarization parameters. A value of $\lambda_{\theta}$ equal to 0, -1 or +1 corresponds to no polarization, longitudinal polarization or transverse polarization scenarios, respectively. The inclusive J/$\psi$ polarization for $p_{\rm T} <$~0.3~GeV/$c$ in the centrality range 70-90$\%$ is found to be compatible with unity within the uncertainties of the measurement, and is also in agreement with similar measurement performed in UPC \cite{ALICE7} (see Fig.\ref{fig34} right) at the same c.o.m energy, in the same rapidity range. According to the measurement in Ref.\cite{ALICE2}, the inclusive J/$\psi$ yield, for $p_{\rm T} <$ 0.3 GeV/$c$ and centrality 70-90$\%$, is dominated by coherently photoproduced J/$\psi$ ($\sim$ 70$\%$). The inclusive J/$\psi$ polarization measurement could then be considered as a good proxy of the coherently photoproduced J/$\psi$ polarization. The result obtained suggests a transverse polarization of coherently photoproduced J/$\psi$ in peripheral Pb--Pb events, in line with expectations from s-channel helicity conservation. 

ALICE has carried out new measurements to probe photoproduced J/$\psi$ in Pb--Pb collisions with nuclear overlap. UPC-like models modified to account for the nuclear overlap are able to describe qualitatively the magnitude of the measured cross section, but fail at describing its rapidity dependence. Stronger constraints on models from comparison to UPC data, and multi-differential measurements towards more central collisions in the future are needed to better understand the impact of the nuclear overlap. The first polarization measurement in peripheral Pb--Pb events gives a consistent picture with earlier results, pointing indeed to a photoproduction origin of these J/$\psi$.


\section*{Acknowledgments}

The author acknowledges the support of the French Agence Nationale de la Recherche (ANR) under reference ANR-22-CE31-0005 MALICE

\end{document}